\documentclass[aps,prl,showpacs,twocolumn,groupedaddress]{revtex4}

\usepackage{graphicx}% Include figure files
\usepackage{dcolumn}% Align table columns on decimal point
\usepackage{bm}% bold math

\begin{document}

%\preprint{}

\title{Characterization of low-energy magnetic excitations in chromium}


\author{H.~Hiraka$^{1,2}$, P.~B\"oni$^3$, K.~Yamada$^2$, S.~Park$^4$, S-.~H.~Lee$^4$
        and G.~Shirane$^1$}
%\thanks{}
\affiliation{ 
   $^1$Physics Department, Brookhaven National Laboratory, Upton, NY 11973, USA\\
$^{2}$Institute for Materials Research, Tohoku University,
Sendai, 980-8577, Japan\\
   $^3$Physik-Department E21, Technische Universit\"at M\"unchen, D-85747 Garching, Germany \\
   $^4$NIST center for Neutron Research,
National Institute of Standards and Technology, Gaithersburg, MD, USA
}


\date{\today}

\begin{abstract}
The low-energy excitations of Cr, i.e. the Fincher-Burke (FB) modes,
have been investigated in the transversely polarized spin-density-wave phase 
by inelastic neutron scattering using a
single-${\bf Q}_{\pm}$ crystal with a propagation
vector ${\bf Q}_{\pm}$ parallel to $[0,0,1]$. 
The constant-momentum-transfer scans show that the energy spectra
consist of two components, namely dispersive 
FB modes and an almost energy-independent cross section.
Most remarkably, we find that the spectrum of the FB modes exhibits
{\it one} peak at 140 K near ${\bf Q} = (0,0,0.98)$ and {\it two}
peaks near ${\bf Q} = (0,0,1.02)$, respectively. 
This is surprising because Cr crystallizes 
in a centro-symmetric bcc structure. 
The asymmetry of those energy spectra decreases 
with increasing temperature.
In addition, the observed magnetic peak intensity
is independent of $\bf Q$ suggesting a transfer of spectral-weight
between the upper and lower FB modes. 
The energy-independent cross section is
localized only between the incommensurate peaks
and develops rapidly with increasing temperature.

\end{abstract}

%\keywords{}

\maketitle

%\section{}
%\subsection{}
%\subsubsection{}

\section{Introduction}

Although chromium exhibits a simple bcc structure and consists
only of a single element the magnetism is very complicated and one
of the most intriguing subjects in condensed matter
physics.~\cite{fawcett88}. Below the N\'{e}el temperature $T_{\rm N}=
311$~K, the magnetic structure exhibits an incommensurate
antiferromagnetic transverse spin-density-wave (TSDW) with the
moments oriented perpendicular to the ordering wavevectors ${\bf
Q}_{\pm}=(2\pi/a)(0, 0, 1\pm\delta)$ ($\delta \simeq 0.045$)
~\cite{Werner67}. Below the spin-flop temperature $T_{\rm sf}=122$~K
the moments arrange along ${\bf Q}_{\pm}$ in a longitudinal
spin-density-wave (LSDW). The pitch of the modulation is
approximately $a/\delta \simeq 20a$ at low temperatures.

\begin{figure}[h]
\centering
     \includegraphics[scale=0.41,clip]{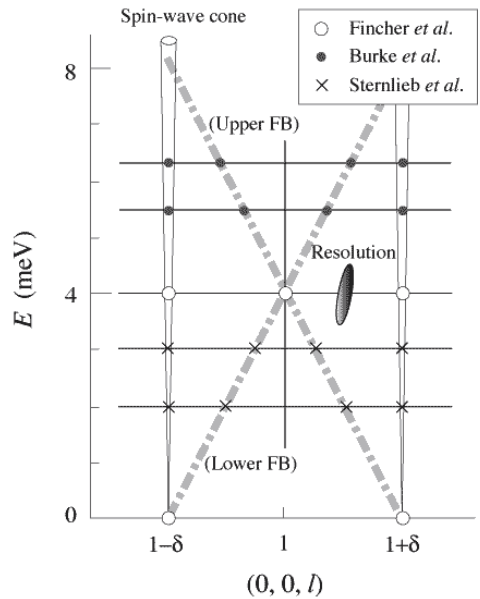} %\vskip 4pt
\caption{Dispersion of magnetic excitations based on the previous
experimental results in the TSDW phase of
Cr. The vertical cones show the dispersion of the high-energy
excitations emerging from the incommensurate 
positions~\cite{fincher81,boeni98}.
The chain lines indicate the FB modes~\cite{fincher79,burke83,sternlieb93} 
and are the
subject of the present investigation. The scan trajectories are
represented by horizontal and vertical lines. The ellipsoid
indicates a typical resolution of the current 
thermal-neutron experiments. }
\label{Fig1}
\end{figure}

The inelastic magnetic cross section shows also a surprisingly
rich behavior (Fig.~\ref{Fig1}). High-velocity excitations (often
called spin waves) emerge from the incommensurate positions ${\bf
Q}_{\pm}$ \cite{fincher81,boeni98}. The excitations exhibit
longitudinal and transverse polarization
\cite{lorenzo94,fukuda96}, the former one being assigned to a
phason mode \cite{fishman96}. In particular, while the
longitudinal mode dominates below an energy transfer $E \simeq 8$
meV, the incommensurate excitations become isotropic above 8 meV.
These modes are present in the LSDW as well as in the TSDW phase.

In addition to the high-energy excitations at ${\bf Q}_{\pm}$,
Fincher {\it et al}.,~\cite{fincher79,fincher81} Burke {\it et
al}.~\cite{burke83} and Sternlieb {\it et al}.~\cite{sternlieb93}
found another magnetic mode in the TSDW phase at low
energies that is located between the incommensurate peaks. The
dispersion relation of this so-called Fincher-Burke (FB) mode was
reported to emanate from the ${\bf Q}_{\pm}$ positions and to
merge into the high-energy excitations near $E = 8$~meV
(Fig.~\ref{Fig1}). The two FB-branches cross each other at 
$E_{\rm c}\sim 4$~meV and ${\bf Q}_{\rm c}=(0,0,1)$. 
(Hereafter, we redefine the two branches as upper and lower
FB mode, respectively, with the boundary energy of $E_{\rm c}$.)
By means of neutron
scattering with polarized neutrons it was shown that the FB modes
are longitudinally polarized~\cite{pynn92,pynn97,boeni98}.

Very recently, Hiraka {\it et al.} re-investigated the FB mode
using unpolarized neutrons and found by means of constant-momentum-transfer 
scans (constant-{\bf Q} scans) a new excitation mode~\cite{hiraka03} 
that emerges from the point $({\bf Q}_{\rm c},E_{\rm c})$
towards a direction transverse to ${\bf Q}_{\pm}$
within the $(h,k,1)$ plane.
Interestingly, the gap-type mode extends with a low velocity
up to at least $h,k = \pm 2\delta$
and shows a decreasing intensity with increasing $h,k$. Using a
cold neutron triple-axis spectrometer with high-resolution B\"oni
{\it et al.} investigated the FB mode in detail~\cite{boeni02}
using constant-$\bf Q$ and 
constant-energy-transfer scans (constant-$E$ scans). 
They pointed out that the FB
mode neither shows a simple linear dispersion nor obeys a simple
spin-wave picture with respect to intensity. In addition, the
intensity contour for $\bf Q_{-} < Q < Q_{+}$ indicated that the FB
mode is asymmetric with respect to ${\bf Q}_{\rm c}$. 
However, it was not clear if this was due to spurious scattering.

In order to obtain a coherent picture of the low-energy
excitations in Cr we have performed a detailed investigation of
the FB modes in the TSDW phase by means of constant-$\bf Q$ scans
using thermal- and cold-neutron spectrometers. Previous
measurements were mostly performed by constant-{\it E} scans
 (Fig.~\ref{Fig1}). However, the intense
high-energy excitations made the quantitative analysis of the weak
response of the FB modes difficult.

The major result of this work is a confirmation of the asymmetry
of the energy spectra with respect to ${\bf Q}_{\rm c}$ near $T_{\rm sf}$
using a different sample and different spectrometers. 
We also observed that the energy spectra become
more symmetric at high temperatures.
In addition,
we show that the FB modes conserve intensity along 
${\bf Q}_{\pm}$ (or $l$),
in contrast to the transverse mode 
that shows a drastic depression of intensity
along $h$ and $k$~\cite{hiraka03}.
The nearly energy-independent component shows a 
strong $T$-dependence 
of the cross section that may be closely related to the
commensurate scattering~\cite{fincher81,fukuda96} or the critical
scattering discussed by Sternlieb {\it et al.}~\cite{sternlieb93}.

%\newpage
\section{Experimental procedure}

For the present experiments we have used the identical single
crystal of $\simeq 4$ cm$^3$ that was previously
investigated~\cite{hiraka03}. The single-$\bf Q$ structure
pointing along $[0,0,1]$ was induced by cooling the sample through
$T_{\rm N}$ in a strong magnetic field yielding a domain population
$\geq 99$~\%. The crystal was mounted inside a closed cycle
refrigerator with the $(h,0,l)$ scattering plane horizontal. The
measurements with thermal and cold neutrons were conducted on the
triple-axis spectrometer TOPAN installed at the JRR-3M research
reactor at the Japan Atomic Energy Research Institute and the
triple-axis spectrometer SPINS at the National Institute for
Standard and Technology, respectively. The $(0,0,2)$ reflection of
pyrolytic graphite was used to monochromate and analyze the
neutron energy. The final neutron energy of TOPAN (SPINS) was
fixed at 14.7~meV (3.7~meV), and the horizontal-collimation
sequence was set to 30'-60'-30'-60' (guide-40'-40'-open) from
before the monochromator to after the analyzer. The energy
resolution of the spectrometers was estimated to be less than
1.2~meV and 0.4~meV, respectively. Higher-order neutrons were
removed by means of a pyrolytic graphite filter and a BeO filter
for TOPAN and SPINS, respectively.

%\newpage
\section{Experimental results}

\begin{figure}[h]
\centering
   \includegraphics[scale=0.7,clip]{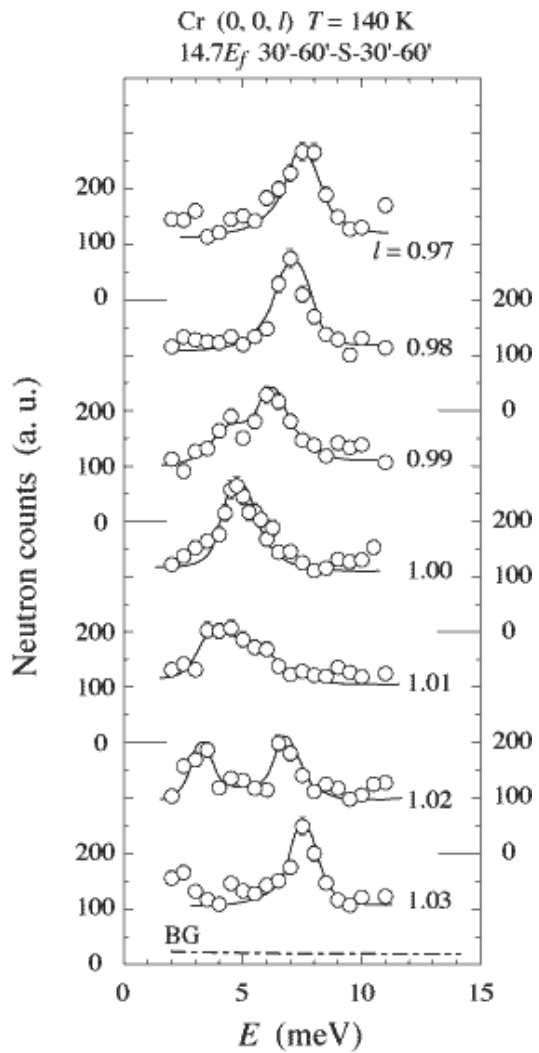} %\vskip 4pt
 \caption{Energy spectra measured along $[0,0,1]$ in the TSDW
 phase. A single peak and a double peak are observed at the symmetric
 $\bf Q$-positions $(0,0,0.98)$ and $(0,0,1.02)$, respectively. The
 solid lines are drawn as a guide to the eyes. 
}
\label{Fig2}
\end{figure}

In Fig.~\ref{Fig2} we show the salient results of the present
work. It is clearly seen that spectra evolve from a single peak at
$(0,0,0.97)$ towards a double-peak structure at $(0,0,1.02)$. The
significance of this result was confirmed by reproducing the scans
under different experimental conditions. 

\begin{figure}[t]
\centering
      \includegraphics[scale=0.7,clip]{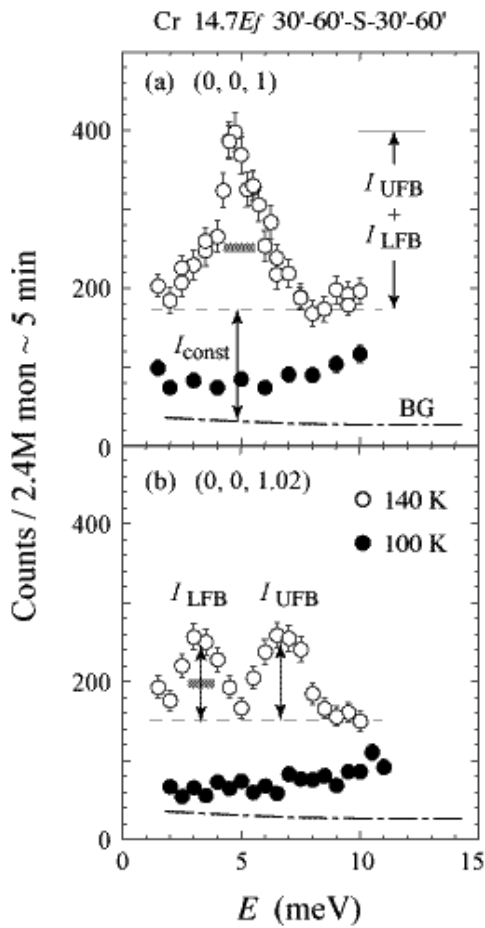} %\vskip 4pt
 \caption{Energy spectra measured above and below $T_{\rm sf} = 122$~K at
 (a) $(0,0,1)$ and (b) $(0,0,1.02)$. 
 $I_{\rm UFB(LFB)}$ represents a peak intensity of the upper (lower) FB mode
 above the $E$-independent cross section $I_{\rm const}$ (broken line)
 at 140 K.
 The chain line indicates the non-magnetic background.
 The energy resolution is shown with horizontal bars.}
\label{Fig3}
\end{figure}

Figure~\ref{Fig3} shows detailed measurements at the commensurate
position $(0,0,1)$ and at the intermediate position $(0,0,1.02)$
for temperatures below and above $T_{\rm sf}$. 
The so-called Fincher peak at $(0,0,1)$~\cite{fincher79} 
is broader than the $E$-resolution.
%indicating a multi-peak structure. 
A broadening of the peaks is also noticed in
the scan at $(0,0,1.02)$. Below $T_{\rm sf}$, the peak structure
disappears completely. At the same time the $E$-independent
magnetic scattering 
decreases too, but it is significantly larger than
the background of $30 \sim 40$ counts/5 min that was measured at
various points in the Brillouin zone.
The behavior of the $E$-independent component is consistent with
``commensurate scattering'' which has already
been observed before by means of constant-$E$ scans using
unpolarized \cite{fukuda96} and polarized neutrons \cite{boeni98}.
%and shown to depend at higher $T$ on $\bf Q$.

\begin{figure}[t]
\centering
    \includegraphics[scale=0.5,clip]{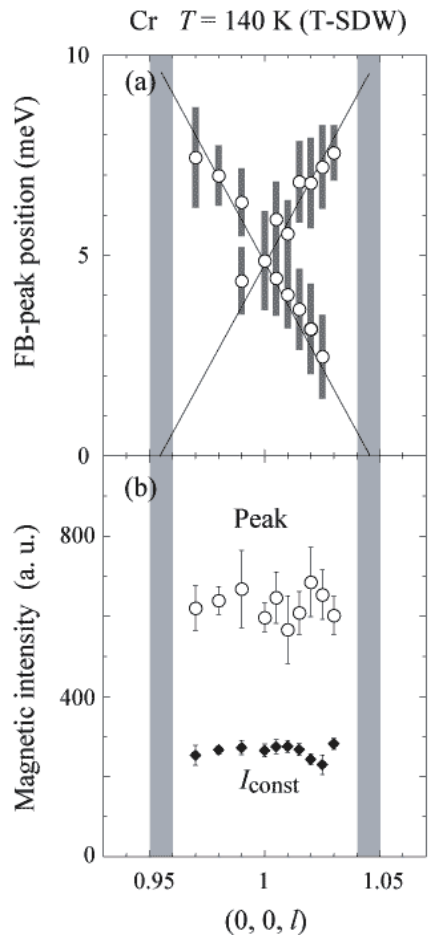} %\vskip 4pt
\caption{$\bf Q$-dependence of (a) the peak position, and (b) the
 observed peak intensity measured at $T$ = 140~K.
The gray region at {\bf Q}$_{\pm}$
 indicates the region where the incommensurate scattering is dominant. 
(a) The peak width is expressed by vertical thick bars. 
In (b), the peak intensity corresponds to a sum of 
$I_{\rm UFB}+I_{\rm LFB}+I_{\rm const}$.
}
\label{Fig4}
\end{figure}

In order to parametrize the data we have fitted the peaks in
Figs.~\ref{Fig2} and \ref{Fig3} after subtraction of the
background to one or two Lorentzians plus a constant, $I_{\rm const}$.
We express peak heights of the former part as $I_{\rm UFB}$
and $I_{\rm LFB}$ for the upper and lower FB mode, respectively (Fig.~\ref{Fig3}).
The latter $I_{\rm const}$ is considered to be
independent of $E$ but depends on $T$. 
Figure~\ref{Fig4}(a) summarizes the $\bf Q$-dependence of 
the energy position of the FB mode. 
The vertical thick bars indicate the width of the peaks
which are, roughly speaking, constant against $\bf Q$.
It is
clearly seen that the dispersion is extremely asymmetric with
respect to ${\bf Q}_{\rm c}$ because no peaks are observed 
at the low-$\bf Q$ side of the lower FB branch.

Figure~\ref{Fig4}(b) shows the peak intensity
defined as a sum of $(I_{\rm UFB}+I_{\rm LFB}+I_{\rm const})$ versus $\bf Q$,
being almost constant between the high-energy incommensurate scattering.
At this stage, no corrections have been made except taking into
account a squared magnetic form factor for Cr$^{2+}$ free ions.
The $E$-independent scattering $I_{\rm const}$ 
contributes about half of the above mentioned peak intensity near $T_{\rm sf}$
and it is insensitive to $l$ also.
Therefore, the intensity of the FB modes ($I_{\rm UFB}+I_{\rm LFB}$)
is independent of $\bf Q$.
%therefore they are not spin-wave excitations. 
At the high-$\bf Q$ side 
a transfer of intensity from the upper to the lower FB branch 
takes place while keeping the total intensity fixed.

\begin{figure}[t]
\centering
     \includegraphics[scale=0.5,clip]{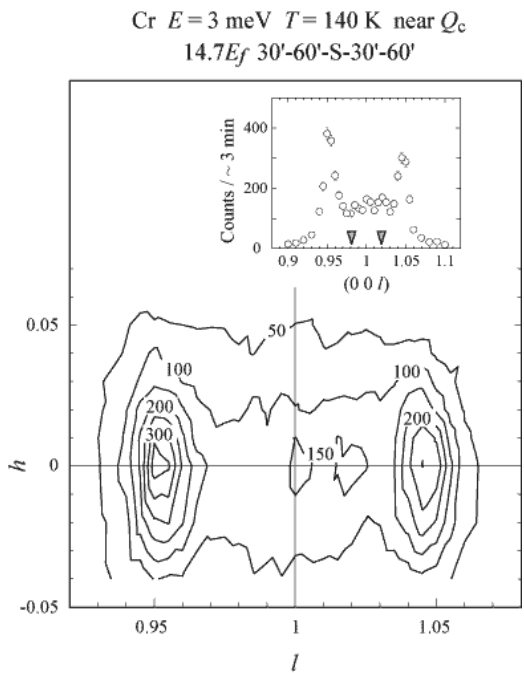} %\vskip 4pt
\caption{Contour map around $(0,0,1)$ for $E = 3$ meV. 
 The intensity around $(0,0,1.02)$ is due to the unbalance of the FB mode
 shown in Fig.~\ref{Fig4}(a). 
 The inset shows a scan keeping the energy transfer fixed at $E = 3$ meV. 
The triangles point towards the $\bf Q$-positions, where FB excitations were
 expected according to Ref.~\cite{sternlieb93}.}
\label{Fig5}
\end{figure}

B\"oni {\it et al.} have already found an asymmetry of the FB mode
for the first time~\cite{boeni02} based on an intensity contour
map constructed from constant-$E$ and constant-$Q$ scans. They
found a blob of scattering at $(0,0,1.02)$ near 3 meV. However,
the quality of the data did not allow to clearly establish the
pronounced (and unexpected) asymmetry in scattering
unambiguously~\cite{boeni-private}. Our new measurements clearly
reproduce the asymmetry between the scattering at $(0,0,0.98)$ and
$(0,0,1.02)$ as shown in Fig.~\ref{Fig5} using a different
single-$\bf Q$ crystal on a different spectrometer with different
resolution.

We notice here once more, that it will be difficult to obtain new
insight from constant-$E$ scans as shown in the inset of
Fig.~\ref{Fig5}, unless the incommensurate scattering at $\bf
Q_\pm$ can be correctly evaluated and convoluted with the
instrumental resolution function. We emphasize that it is the
constant-$\bf Q$ technique that allows the unambiguous
determination of dispersion relations and not the constant-$E$
technique that leads often to wrong conclusions as proven in the
past.

\begin{figure}[t]
\centering
     \includegraphics[scale=0.7,clip]{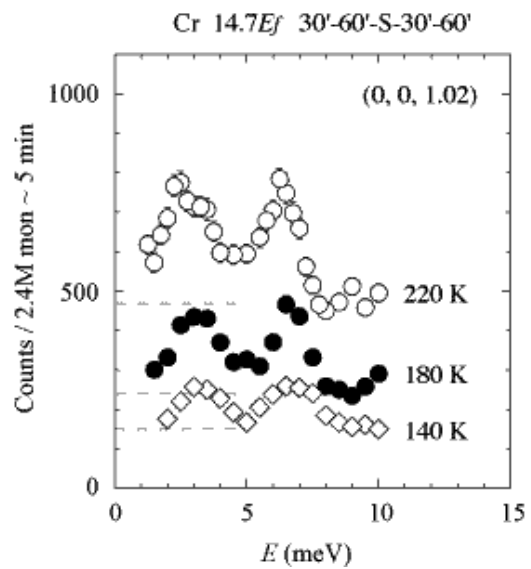} %\vskip 4pt
\caption{Thermal evolution of energy spectra at $(0,0,1.02)$ from
 140~K to 220~K. The horizontal broken lines
show the magnetic scattering intensity presumed at each temperature.
The non-magnetic background level is estimated to be $<40$ coutns /5 min.
}
\label{Fig6}
\end{figure}

\begin{figure}[t]
\centering
     \includegraphics[scale=0.7,clip]{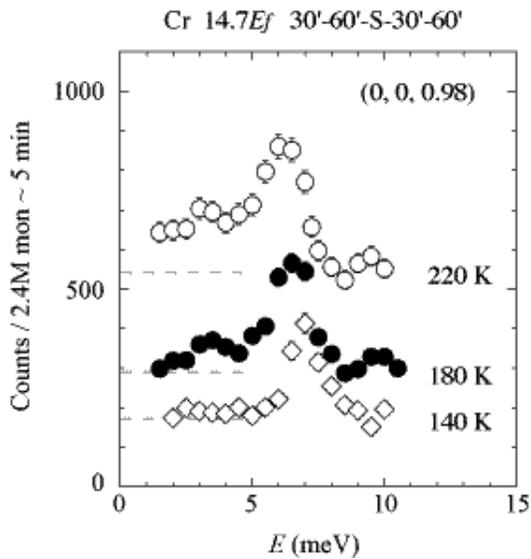} %\vskip 4pt
\caption{Thermal evolution of energy spectra at $(0,0,0.98)$.
 Note the development of the ``missing'' peak near 3 meV with increasing $T$. 
}
\label{Fig7}
\end{figure}

Figures~\ref{Fig6} and \ref{Fig7} show the evolution of the
inelastic intensity at $(0,0,1.02)$ and $(0,0,0.98)$ with $T$,
respectively. 
The magnetic intensity increases much faster than expected on
the basis of the thermal population factor, $\langle n+1 \rangle =
1/(1-\exp(-E/k_{\rm B}T))$, indicating that the scattering
is not due to spin waves \cite{clementyev04}. 
In particular,
a remarkable growth is seen in $I_{\rm const}$
as shown with shades.
The increase is also not due to conventional critical scattering near
$T_{\rm N}$ because the measurements were performed far away from
$T_{\rm N} = 311$ K, i.e. at reduced temperatures 
$t = 1-(T/T_{\rm N})$ of 0.58 and 0.71 for 180~K and 220~K, respectively. 
We point out that
with increasing $T$ the ``missing'' peak develops at $(0,0,0.98)$
near 3 meV (Fig.~\ref{Fig7}), i.e. the spectrum becomes more
symmetric with increasing $T$ and approaches the cross section at
$(0,0,1.02)$ (Fig.~\ref{Fig6}).

\begin{figure}[t]
\centering
     \includegraphics[scale=0.7,clip]{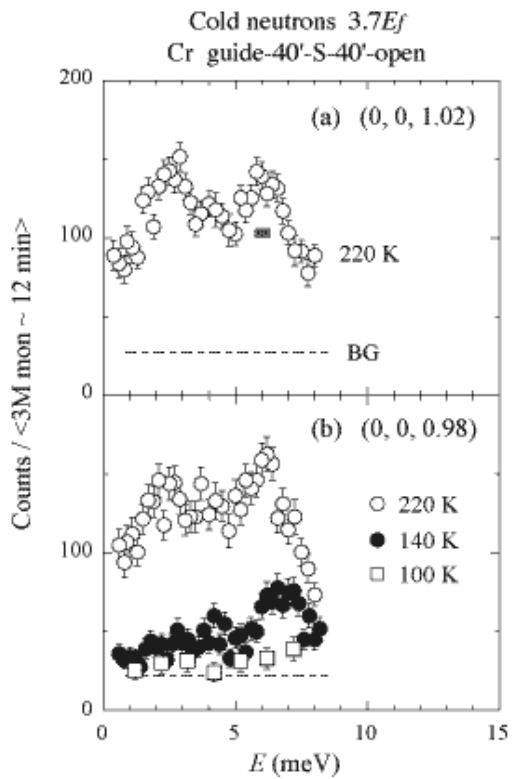} %\vskip 4pt
\caption{Cold-neutron data of constant-$\bf Q$ scans
 at the symmetrical positions (a) $(0,0,1.02)$ and (b)
 $(0,0,0.98)$. The energy resolution of $\Delta E \leq 0.4$~meV (FWHM)
 is shown by the horizontal bar in (a).
 The background level was estimated from data at two points
 $(0,0,0.9)$ and $(0,0,1.1)$.}
\label{Fig9}
\end{figure}

Figure~\ref{Fig9} shows (i) once more the asymmetry of the
scattering with respect to ${\bf Q}_{\rm c}$ 
and (ii) the reduction of unbalance for two peaks 
at $(0,0,0.98)$ with increasing $T$. 
These high-resolution measurements ($\Delta E \le 0.4$ meV in FWHM) 
confirm the results from Fig.~\ref{Fig7} ($\Delta E \le
1.2$ meV) that were performed with thermal neutrons. Due to the
very different resolution conditions the asymmetry of the
scattering is less pronounced with cold neutrons.

%\newpage
\section{Discussions}

We have shown that the FB modes can be measured more precisely by
means of constant-$\bf Q$ scans than by means of constant-$E$
scans. As a result, it was possible to clearly identify an
asymmetry of the magnetic cross sections with respect to the
commensurate position ${\bf Q}_{\rm c}$.
The asymmetry decreases with increasing $T$. We have proven by using very
different experimental conditions and different spectrometers that
the asymmetry is not due to a resolution effect. The asymmetry is
also not an artifact of the sample because similar results were
obtained using another Cr single crystal and another spectrometer
\cite{boeni02}. It might be interesting to investigate the
asymmetry of the scattering in other Brillouin zones.

Our results also indicate that the FB modes are not spin-wave
modes because the peak intensity is independent of
$\bf Q$ (Fig.~\ref{Fig4}(a)). 
Moreover, we observe a transfer of intensity between the
lower and upper branches of the excitations. Finally, the
intensity of the modes increases significantly faster with
increasing $T$ than given by $\langle n+1 \rangle$,
which is in
agreement with previous works \cite{fincher81,clementyev04}. 
One may speculate that we have to deal with a thermally activated
process. 
Another possible explanation may be reached by considering
mode-mode coupling between the upper and lower FB branches~\cite{harada71}.

The present results strongly suggest the existence of magnetic 
scatterting $I_{\rm const}$ that develops
around the commensurate ${\bf Q}_{\rm c}$ position. 
Its intensity increases also significantly faster
than given by $\langle n+1 \rangle$ and finally dominates the
inelastic scattering between $\bf Q_{\pm}$ in agreement with
previous constant-$E$ works \cite{fincher81,fukuda96}. 
$I_{\rm const}$ should contribute to
the significant critical scattering near $T_{\rm N}$ \cite{fincher81}.
In contrast to the FB modes this scattering is polarization
independent \cite{boeni98} and exists already in the LSDW phase
below the spin-flop transition. 
The energy-independent feature of $I_{\rm const}$ is
also present transverse to $\bf Q_\pm$ but drastically decreases with
increasing $h$ and $k$ \cite{hiraka03}. 
We do not know the origin yet.
One may speculate that the
FB modes as well as the $E$-independent scattering may be explained by
the complicated shape of the Fermi surface that gives rise to many
different modes. We mention in particular that already 
a three-band model gives rise to at least 25 magnetic modes
\cite{fishman96}.

%\newpage
\section{Conclusion}

We have reinvestigated the low-energy FB excitations of Cr in the
TSDW phase by making extensive use of constant-$\bf Q$ scans. We
have observed an asymmetry in the inelastic scattering near $T_{\rm sf}$
that is at variance with the simple centro-symmetric bcc structure of Cr. 
The asymmetry of energy spectra decreases with raising $T$.
The observed peak intensity of the FB modes satisfies a sum rule
with regard to {\bf Q}.
The intensity of the dispersive excitations 
and the $E$-independent
scattering increase with increasing $T$ much faster than expected
according to the temperature factor $\langle n+1 \rangle$.
Therefore, the FB modes are not spin-wave modes. 
We expect that
the present results challenge the development of a theory to
explain these results and resolve the puzzle of the magnetic
excitations in Cr.

%\newpage
\begin{acknowledgments}


We thank Y.~Endoh, R.~S.~Fishman, S.~A.~Werner and B.~J.~Sternlieb
for helpful discussions. Present research was supported by the
U.S.-Japan Cooperative Neutron-Scattering Program. Work at Tohoku
University was supported by the Ministry of Monbu-Kagaku-shou of
Japan. Work at Brookhaven was supported by the Division of
Material Sciences, U.S. Department of Energy under contract
DE-AC02-76CH00016. Work at SPINS was based upon activities
supported by the NSF under DMR-9986442.

\end{acknowledgments}

%\newpage

\bibliography{basename of .bib file}

\end{document}